\pdfoutput=1
\documentclass{article}
\usepackage{graphicx, indentfirst, hyperref, natbib}
\usepackage[a4paper, margin = 3cm]{geometry}
\makeatletter
\let\mailmark\@fnsymbol
\makeatother

\newcommand*{\cf}{\emph{cf.}}
\newcommand*{\eg}{\emph{eg.}}
\newcommand*{\etc}{\emph{etc}}
\newcommand*{\prog}[1]{\emph{#1}}
\newcommand*{\figref}[1]{Figure \ref{fig:#1}}
\newcommand*{\secref}[1]{Section \ref{sec:#1}}
\let\thxmark\textsuperscript
\let\cite\citep
\let\citeasnoun\citet
\begin{document}
\title{%
	A versatile framework for attitude tuning of\\
	beamlines at advanced light sources%
}
\author{%
	Peng-Cheng Li\thxmark{1,2,3,\mailmark{1}},
	Xiao-Xue Bi\thxmark{2,\mailmark{1}},
	Zhen Zhang\thxmark{1,3}, Xiao-Bao Deng\thxmark{2},\\
	Chun Li\thxmark{2}, Li-Wen Wang\thxmark{2},
	Gong-Fa Liu\thxmark{1,3}, Yi Zhang\thxmark{2,3},
	Ai-Yu Zhou\thxmark{2}, Yu Liu\thxmark{2,\mailmark{2}}%
}
\date{}
\maketitle
\begingroup
\renewcommand{\thefootnote}{\fnsymbol{footnote}}
\footnotetext[1]{\ Li and Bi contributed equally to this paper.}
\footnotetext[2]{\ Correspondence e-mail: \texttt{liuyu91@ihep.ac.cn}.}
\endgroup
\footnotetext[1]{\ %
	National Synchrotron Radiation Laboratory,
	University of Science and Technology of China,
	Hefei, Anhui 230029, People's Republic of China.%
}
\footnotetext[2]{\ %
	Institute of High Energy Physics, Chinese Academy of Sciences,
	Beijing 100049, People's Republic of China.%
}
\footnotetext[3]{\ %
	University of Chinese Academy of Sciences,
	Beijing 100049, People's Republic of China.%
}

\section*{Synopsis}

Keywords: beam focusing, sample alignment,
\prog{Mamba}, virtual beamline, software architecture.

A versatile \prog{Mamba}-based software framework for automated attitude
tuning of beamlines is reported, which in our assessment is able to cover
a majority of attitude-tuning (beam focusing, sample alignment \etc)
needs in a simple and maintainable way.  Apart from a few real-world
examples at HEPS and BSRF, also presented is a virtual-beamline
mechanism based on easily customisable simulated detectors and motors.

\section*{Abstract}

Aside from regular beamline experiments at light sources, the preparation steps
before these experiments are also worth systematic consideration in terms of
automation; a representative category in these steps is attitude tuning, which
typically appears in names like beam focusing, sample alignment \etc.  With
the goal of saving time and manpower in both writing and using in mind, a
\prog{Mamba}-based attitude-tuning framework is created.  It supports flexible
input/output ports, easy integration of diverse evaluation functions, and
free selection of optimisation algorithms; with the help from \prog{Mamba}'s
infrastructure, machine learning (ML) and artificial intelligence (AI)
technologies can also be readily integrated.  The tuning of a polycapillary
lens and of an X-ray emission spectrometer are given as examples for the
general use of this framework, featuring powerful command-line interfaces
(CLIs) and friendly graphical user interfaces (GUIs) that allow comfortable
human-in-the-loop control.  The tuning of a Raman spectrometer demonstrates
more specialised use of the framework with customised optimisation algorithms.
With similar applications in mind, our framework is estimated to be capable
of fulfilling a majority of attitude-tuning needs.  Also reported is a
virtual-beamline mechanism based on easily customisable simulated detectors and
motors, which facilitates both testing for developers and training for users.

\section{Introduction}\label{sec:intro}

In beamline experiments, apart from the main scan steps (``counting'', step
scans or fly scans, also including their data processing), the preparation
steps before them can also be of considerable complexity, and therefore be
of particular interest in terms of automation.  A representative category
in these steps is beam focusing \cite{nash2022, takeo2020, hong2021, xi2017}
and sample alignment \cite{robertson2015, zhang2023} \etc.  At HEPS, the High
Energy Photon Source \cite{xu2023} and BSRF, the Beijing Synchrotron Radiation
Facility, we refer to these work as \emph{attitude tuning}, borrowing the
term ``attitude'' from aerospace engineering (the term ``configuration'' could
also be considered, but would be very ambiguous).  In our eyes, the essence
of attitude tuning is the optimisation of certain objective parameter(s)
by manual or automated tuning of the corresponding attitude parameters:
\eg\ the automatic focusing of cameras is the automatic optimisation of some
image definition functions by tuning parameters like focal lengths.  From
the information we have collected from BSRF, HEPS and our colleagues' visits
to other facilities, we find attitude tuning a ubiquitous requirement at these
facilities; but given the background above, we also find that despite the
diversity of devices involved in these requirements, a majority of them are
fairly simple peak finding, albeit often in $>$1D configuration spaces.

As HEPS and other advanced light sources have small light spots and high
brightness, traditional scan-based methods for attitude tuning could not
only waste a lot of time (especially in $>$1D tuning applications), but also
potentially result in more radiation damages to samples.  Also considering
that attitude-tuning requirements need to be implemented for the 15 beamlines
in Phase I of HEPS, it is imperative for us to create an efficient and
unified software framework for attitude tuning.  Fortunately, \prog{Mamba}
\cite{liu2022, dong2022}, the \prog{Bluesky}-based \cite{allan2019} software
environment created for beamline experiments at HEPS, was also designed with
attitude tuning and other preparation steps in mind since the very beginning.
Based on it, we created a versatile framework for attitude tuning, where
peak finding can be done in a simple, consistent and maintainable way.
With the help from \prog{Mamba}'s infrastructure and the rich selection
of third-party Python libraries, more complex tuning, including those
that require machine learning (ML) and/or artificial intelligence (AI)
technologies \cite{rebuffi2023, zhang2024}, can also be implemented.
The code for our framework is available as a part of the open-source
edition of \prog{Mamba} at \url{https://codeberg.org/CasperVector/mamba-ose}.

\section{Architecture of the attitude-tuning framework}\label{sec:arch}

\begin{figure}[htbp]\centering
\includegraphics[width = 0.5\textwidth]{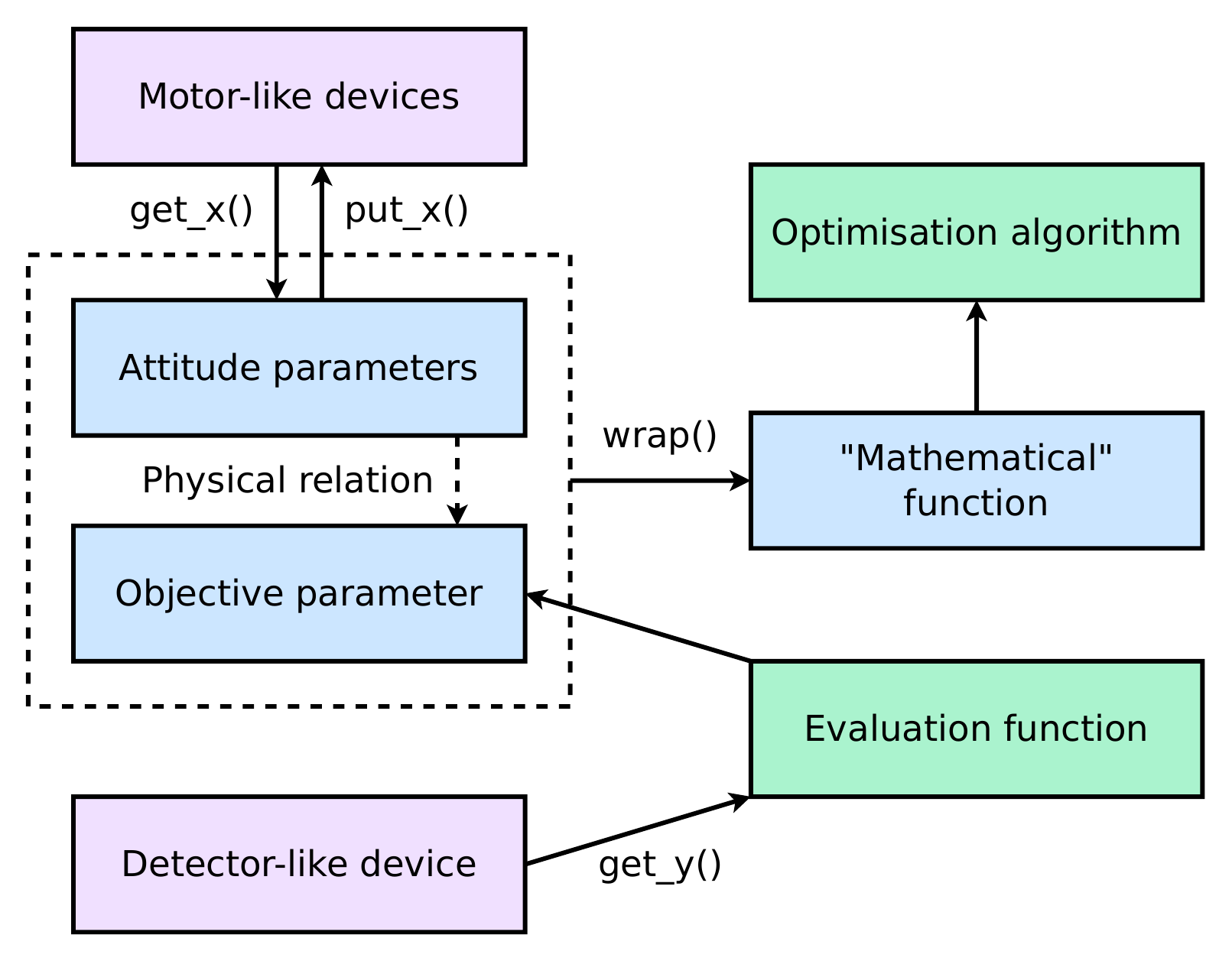}
\caption{%
	Architecture of our attitude-tuning framework based on \texttt{AttiOptim}.%
}\label{fig:atti-arch}
\end{figure}

As has been noted in \secref{intro}, we treat attitude tuning as a matter
of numerical optimisation; therefore the architecture of a general-purpose
attitude-tuning framework (\figref{atti-arch}) will inevitably include
some attitude parameters, some objective parameter(s), and some optimisation
algorithm.  Moreover, since the objective parameter(s) are actually obtained
by physical measurement instead of purely mathematical computation, the
architecture also needs to include detectors, motors and evaluation
functions which convert raw data from detectors into objective values.
Given these architectural elements, we implemented the \verb|AttiOptim|
class which cooperates with \prog{Bluesky}'s interfaces for motors and
detectors, as well as optimisation libraries like \verb|scipy.optimize|.
The functions \verb|get_x()|, \verb|put_x()| and \verb|get_y()| in this
class deal with motors and detectors by interacting with their \prog{Bluesky}
encapsulations.  The function \verb|wrap()| combines specified processing/%
evaluation functions with \verb|put_x()| and \verb|get_y()| into functions
with a seemingly purely mathematical signature, which are wanted by
libraries like \verb|scipy.optimize| (\cf\ \figref{lib-4w1b}).

In \secref{capixes}, we will see real-world application of our framework,
using \verb|AttiOptim|, on a polycapillary lens and an X-ray emission
spectrometer: the former is more general, with a straightforward processing/%
evaluation function; the latter is more specialised, and demonstrates the use
of a much more complex evaluation function.  Then in \secref{raman}, we will
see even more specialised application of our framework on a Raman spectrometer,
where the ``optimisation algorithm'' interface (\cf\ also \secref{discuss}) is
``abused'' to do parallelised peak finding of multiple objective parameters,
as well as some task that is even not numerical optimisation at all.  By
customising \verb|get_x()|, \verb|put_x()| and \verb|get_y()|, it is also
possible to manipulate general motor-like devices, and possible to achieve
effects like using the results from some kind of scan \cite[\eg\ tomography
or ptychography, \cf][]{zhang2023} as the raw data $y$ for each position
$x$, where the objective value $z$ is computed from.  With the help from
\prog{Mamba}'s infrastructure, ML/AI technologies can also be easily integrated
in the evaluation function and/or the optimisation algorithm.  On a deeper
level, following \citeasnoun{hutter2020}'s treatment of AI as a matter of
information compression, the pursuit of minimalist solutions for problems
throughout our work in beamline control and in experiment software
\cite{li2024} may be also seen as, in essence, a pursuit of \emph{human
intelligence} which may enlighten its AI counterparts in the same fields.

\section{General attitude tuning: a polycapillary lens and an XES spectrometer}
\label{sec:capixes}

\begin{figure}[htbp]\centering
\includegraphics[width = 0.8\textwidth]{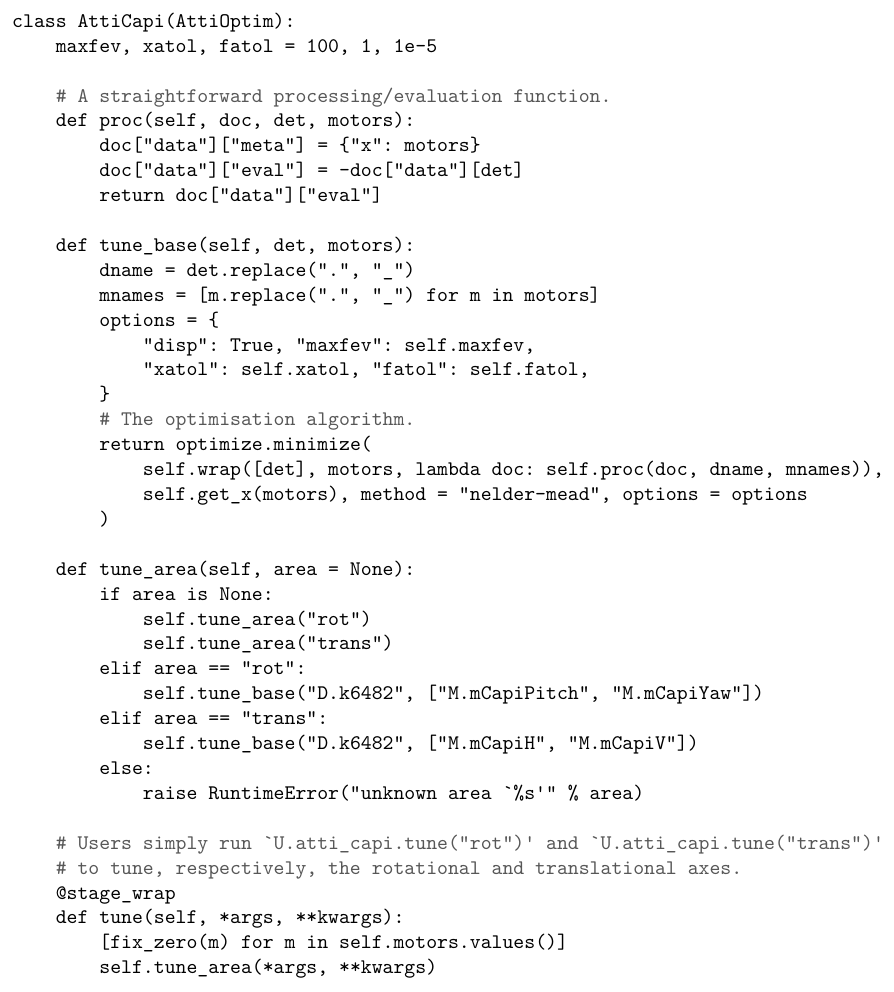}
\caption{%
	Notable fragments of \texttt{lib\string_4w1b.py} with brief usage notes.%
}\label{fig:lib-4w1b}
\end{figure}

\begin{figure}[htbp]\centering
\includegraphics[width = 0.9\textwidth]{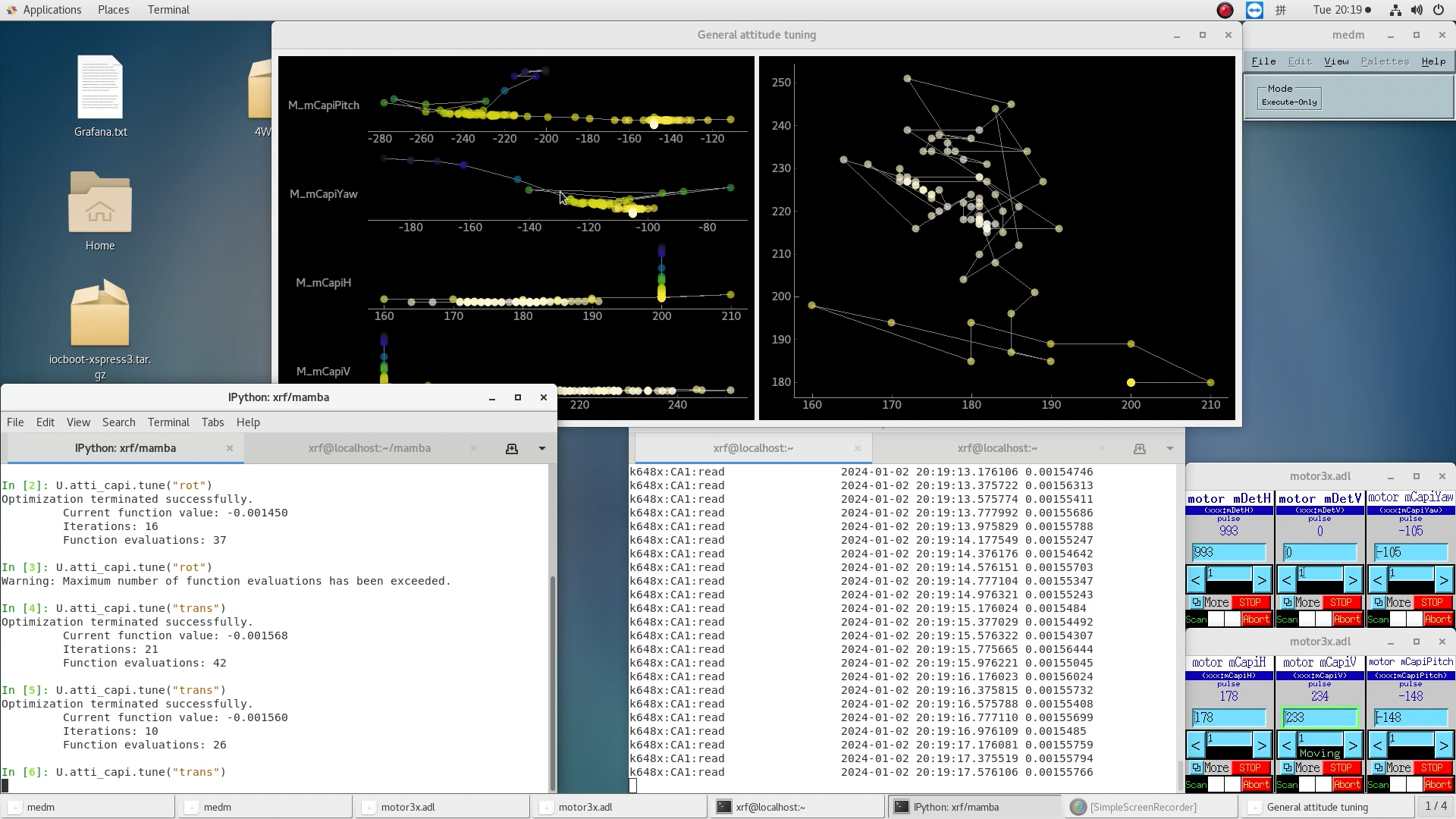}
\caption{Attitude tuning of the polycapillary lens at 4W1B of BSRF.}
\label{fig:capi-gui}
\end{figure}

Our first example is about the polycapillary lens at the 4W1B beamline
of BSRF, which has 4 motorised degrees of freedom: a pitch/tilt angle
(\verb|M.mCapiPitch|), a yaw/pan angle (\verb|M.mCapiYaw|), a horizontal
shift (\verb|M.mCapiH|) and a vertical shift (\verb|M.mCapiV|).  Based on
prior experience, they are split into 2 groups: the rotational parameters
are the coarse tuning parameters, and the translational parameters are
the fine tuning parameters; the former usually need to be tuned before
the latter.  The goodness of the lens' attitude is determined by the
readings from a Keithley 6482 picoammeter (\verb|D.k6482|), which is
connected to a photodiode temporarily placed next to the lens when
attitude tuning is performed.  The code used for this tuning based on
our framework, \verb|init_capi.py| and \verb|lib_4w1b.py|, are available
from the supplementary materials; the former mainly does \prog{Bluesky}
encapsulation of the devices involved, while the latter (with notable
fragments and brief usage notes in \figref{lib-4w1b}) contains the
tuning logic.  Online visual feedback of the tuning process can be done
with our general-purpose graphical user interface (GUI) for attitude
tuning, \verb|mamba.attitude.capi_frontend| (\figref{capi-gui}).  A more
feature-complete version of the GUI is shown in \figref{sim-gui}(a), which
most notably allows for the manual selection of axes (called $x$ and $y$
therein) for the 2D visualisation on the right pane; otherwise the GUI
would automatically select the axes based on the latest data update.

Attitude tuning for the polycapillary lens at 4W1B of BSRF needs to be done
roughly once per day during normal operation of the beamline.  It used to
be done by manual trial and error, each time costing around half an hour;
with our attitude-tuning framework, the process has been greatly accelerated
and simplified, costing just a few minutes each time and demanding much less
manual intervention.  From the information we have collected, attitude-tuning
requirements structurally similar to the above are ubiquitous at HEPS, BSRF
and other facilities, even if the devices and attitude parameters may differ
greatly across beamlines.  For these reasons, we believe our framework
is general enough to handle this kind of application scenarios, saving
great amounts of time and energy while only requiring very modest amounts of
programming.  We would also like to note that we are aware of the potential
needs to tune multiple groups of attitude parameters, where not all groups
are tuned according to the same objective parameter.  In response, our
framework is designed to be capable of \emph{multi-objective tuning}:
an example for it is given in the files \verb|docs/init_capi.py| and
\verb|docs/lib_capi.py| from the open-source edition of \prog{Mamba}; in
the general-purpose visualisation GUI (\figref{sim-gui}a), the objective
parameter to visualise can be selected from the $z$ axis menu.

\begin{figure}[htbp]\centering
\includegraphics[width = 0.9\textwidth]{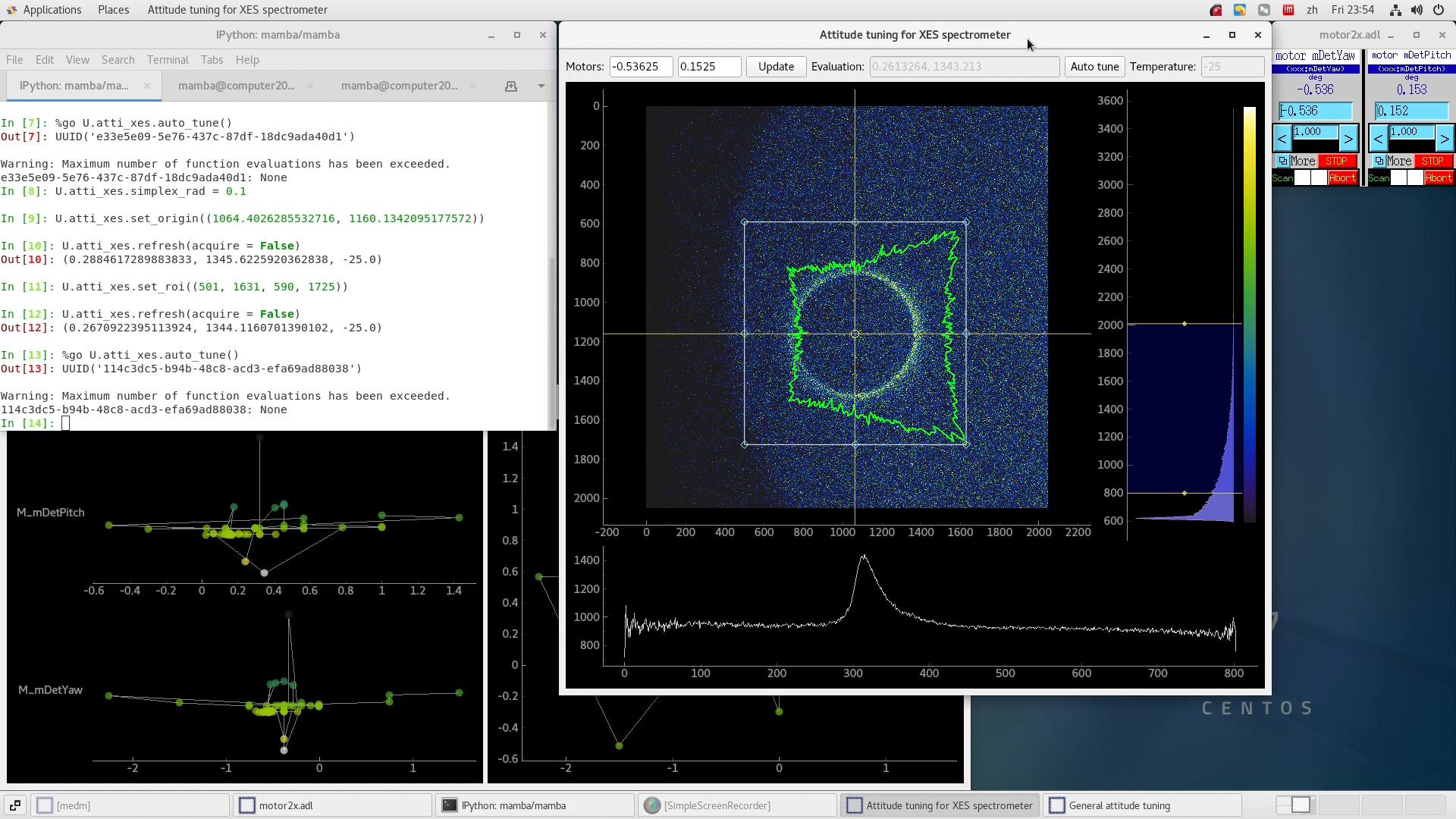}
\caption{Attitude tuning of an XES spectrometer at 4W1B of BSRF.}
\label{fig:xes-gui}
\end{figure}

Another example for our attitude-tuning framework is about a full-cylindrical
von Hamos spectrometer \cite{guo2023} for X-ray emission spectroscopy
(XES), currently used at 4W1B of BSRF; its use at the high pressure beamline
(B6), the X-ray absorption spectroscopy beamline (B8) and possibly other
beamlines at HEPS is also planned.  Bragg reflection from the analyser of
the spectrometer produces circular patterns in images acquired from the
detector (\figref{xes-gui}); the pitch/tilt and yaw/pan angles are tuned
to optimise the shape of the circle, so that a sharpest peak is obtained
on the radial distribution curve, shown on the lower pane of the main
window in the figure.  More precisely, the half-maximum region of interest
(HM-ROI) is computed for the radial distribution, then the mean intensity
in this ROI is used as the objective parameter.  From the description
above it can already be seen that for the tuning of this spectrometer,
the processing/evaluation function needed must have a non-trivial
complexity.  But in addition to this, another main source of complexity
is the selection of centre (origin) of the circles, where inappropriately
selected origins can result in distorted peaks and unoptimal attitudes.

Noticing that automated origin selection is a computationally expensive
process which however only needs to be done after significant changes
to the attitude, we decided to use a \emph{human-in-the-loop} approach
for it.  An ``Auto origin'' button is provided in the GUI (\cf\ the more
feature-complete version in \figref{sim-gui}a) for automated fine tuning
of the origin, whose coarse selection can be done manually by dragging
the origin point shown in the GUI.  In the light of the analyses above, we
wrote the specialised attitude-tuning program for this XES spectrometer as
\verb|mamba.attitude.xes_backend| and \verb|mamba.attitude.xes_frontend|,
where the latter is the main GUI shown in the figures above.  Considering
that radial distributions and raw images are the main data wanted by users
in normal data acquisition (``counting'') after attitude tuning has been
done, this program is also written with normal counting in mind: the GUI
provides features wanted by users, like setting acquisition time and saving
experiment data.  As attitude tuning is normally done by beamline staffs,
and the tuning process (except for origin selection) is already convenient
enough for them on the command line, there is no GUI button for it.
Other than the general-purpose \verb|mamba.attitude.capi_frontend| (as
is with the polycapillary lens above), visualisations in the specialised
GUI will also be automatically updated after each move in the tuning process.

\section{Specialised attitude tuning: an XRS spectrometer}\label{sec:raman}

\begin{figure}[htbp]\centering
\includegraphics[width = 0.95\textwidth]{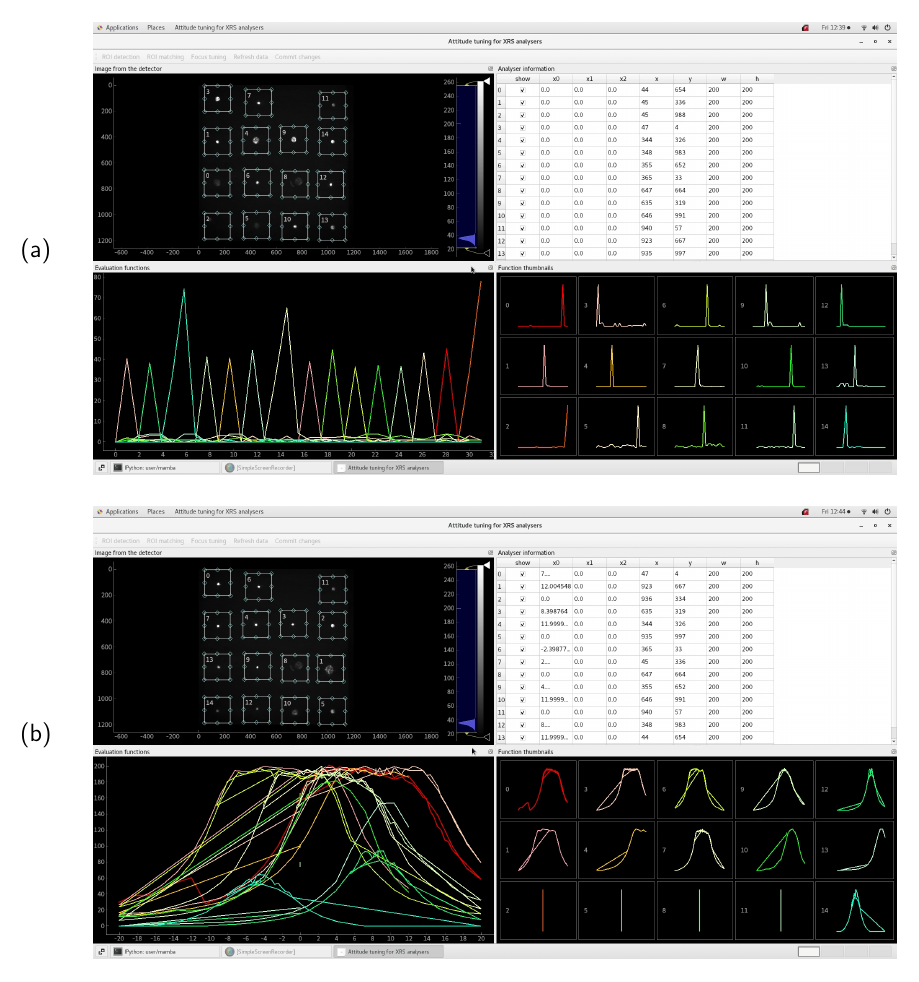}
\caption{%
	GUI of our XRS attitude-tuning program, doing (a) ROI matching and (b)
	focus tuning; the pictures are obtained from a laser-based simulation
	of what would eventually be done with X-ray at HEPS.  Motor motion and
	ROI editing can be done on the top-right pane; ROIs can also be modified
	with mouse operations on ROI rectangles on the top-left pane, and/or with
	drag-and-drop operations of table rows on the top-right pane.  Automated
	tuning is currently only implemented for a single analyser module, while
	parallelised tuning of multiple modules will be implemented in the future.%
}\label{fig:raman-gui}
\end{figure}

Among the instruments at the hard X-ray high-resolution spectroscopy
beamline (B5) of HEPS, currently under active construction, is an X-ray
Raman scattering (XRS) spectrometer.  Structurally similar to the spectrometer
in \cite{huotari2017}, the XRS spectrometer at B5 of HEPS has 6 analyser
modules, each containing one detector and a 3$\times$5 array of analysers;
each analyser has three motorised degrees of freedom -- one longitudinal
shift and two latitudinal angles.  As is shown in \figref{raman-gui},
latitudinal tuning ($x_1$ and $x_2$ for each analyser) moves the X-ray
spots around on detectors, while longitudinal tuning ($x_0$ for each
analyser) focuses the spots.  After the spots are properly distributed
and each of them is correctly assigned to the corresponding analyser,
focus tuning is performed; all these steps can be done in the GUI
of our specialised attitude-tuning program for this spectrometer,
\verb|mamba.attitude.raman_backend| and \verb|mamba.attitude.raman_frontend|.
Spot distribution is done manually with visual aid from the GUI, spot
assignment (ROI detection followed by ROI matching) is automated, while
focus tuning is automated and parallelised.  Due to the use of multiplexers
for motion controllers \cite{li2024}, the motors in each analyser module
are separated into multiple groups, where two motors in the same group
cannot be moved at the same time.  Consequently, in our program the
parallelisation of focus tuning is done in multiple passes: the
first motor in every longitudinal group, then the second, \etc.

Focus tuning of the XRS spectrometer at B5 of HEPS is based on a 2D
generalisation of the HM-ROI mean in \secref{capixes} as the objective
parameter.  Other than application on spectrometers in this paper, similar
evaluation functions have also been used in a few other scenarios, \eg\ the
standalone X-ray beam position monitor (XBPM) program in \cite{li2024} based
on images from area detectors.  The optimisation is implemented by multiple
calls to \verb|max_parascan()|, a function that optimises, in parallel, the
objective parameter for each ROI in the list of ROIs passed to this function.
This function's signature is similar to \eg\ \verb|scipy.optimize.minimize()|,
except that its objective function produces an 1D numerical array instead
of a 0D number.  Internally, it first does a coarse inner-product scan
(\cf\ \verb|scan()| from \prog{Bluesky}) for elements of the input
parameter, then does a fine inner-product scan of the input elements
in the reverse direction, and finally sets the input elements to the
peak positions (\cf\ \figref{raman-hyst} for the details).  ROI matching
is implemented with \verb|perm_diffmax()|, a function with a signature
similar to \verb|max_parascan()|; its ``objective function'' computes all
X-ray spots' barycentres inside their HM-ROIs (which are also used in the
XBPM program mentioned above), as well the distance from each barycentre to
a corresponding reference position.  For each analyser, \verb|perm_diffmax()|
sets the reference positions to the initial positions of the barycentres;
it then changes the latitudinal parameters of this analyser in small steps
until a spot/ROI stands out with a significantly larger distance than those
of all the rest.  This outstanding ROI is assigned to the current analyser,
and the latitudinal parameters are reset to the original values; the reference
positions are updated to counteract potential motor backlashes, and then
\verb|perm_diffmax()| moves on to the next analyser.  \verb|perm_diffmax()|
finally returns the mapping table from ROIs to analysers, which is
applied in the command-line backend and GUI frontend of our program.

\begin{figure}[htbp]\centering
\includegraphics[width = 0.8\textwidth]{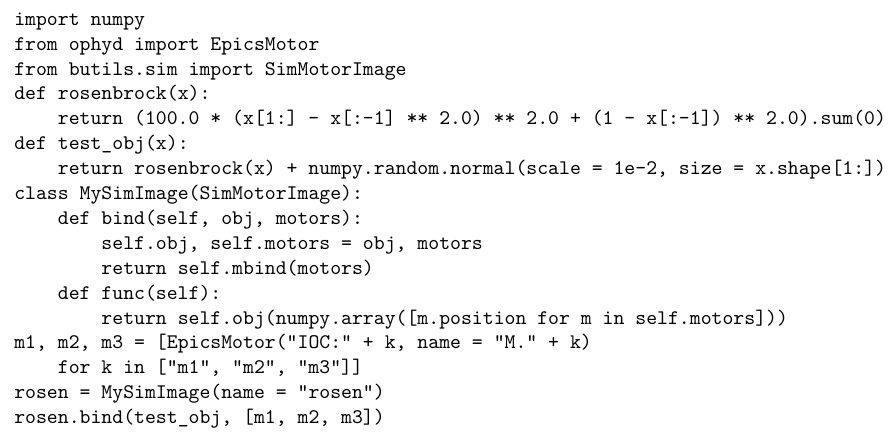}
\caption{%
	A virtual device connected to 3 \prog{motorMotorSim}-based
	motors, based on the Rosenbrock function widely used to
	test optimisation algorithms, perturbed by a random noise.%
}\label{fig:sim-rosen}
\end{figure}

\begin{figure}[htbp]\centering
\includegraphics[width = 0.95\textwidth]{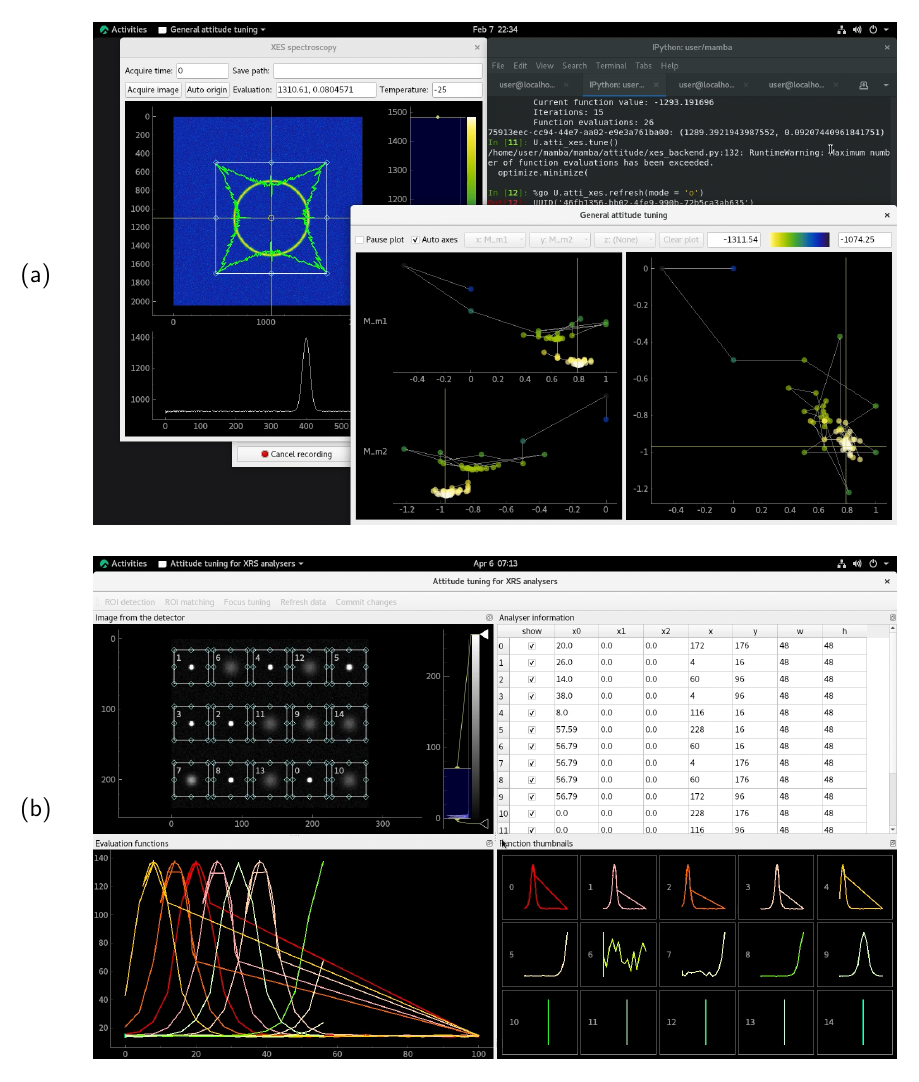}
\caption{%
	Our (a) XES and (b) XRS attitude-tuning
	programs running with virtual beamlines.%
}\label{fig:sim-gui}
\end{figure}

In the end of this section, we briefly introduce the virtual-beamline mechanism
which we now use extensively both in attitude tuning and in other applications.
First, based on the \prog{motorMotorSim} and \prog{ADSimDetector} modules in
\prog{EPICS}, we are already able to perform many kinds of simulations, also
including those independent of \prog{Bluesky}, \eg\ the Python IOC for motor
multiplexers in \cite{li2024}.  We note that the former is especially useful
because of its realistic simulation of motor speeds, soft limits \etc.  Second,
we also created the \verb|butils.sim| module for \prog{Mamba}, which provides
useful and easy-to-use simulation device classes that expose interfaces
similar to those of \prog{Bluesky}'s classes for real devices.  For example,
the \verb|SimMotorImage| class (\figref{sim-rosen}) implements a virtual device
that binds to a simulation function and some motor-like device object(s),
whether real motors, simulated motors like those based on \prog{motorMotorSim},
or even things like energies of monochromators; it produces readings according
to the device positions and the simulation function.  Based on the mechanisms
above, we are able to create virtual beamlines to test programs and train
staffs/users, saving lots of beamtime and allowing for the development of
software before the required instruments are fully ready.  For instance, with
virtual beamlines we can test our attitude-tuning programs (\figref{sim-gui};
all the simulation code is available in the \verb|docs| subdirectory of
the open-source edition of \prog{Mamba}) extensively before their tests on
real hardware, and often only need minor tweaks/fixes in the latter tests.

\section{Outlook and discussion}\label{sec:discuss}

With the real-world examples in Sections \ref{sec:capixes}--\ref{sec:raman},
and noticing the architectural versatility of our attitude-tuning framework
(\cf\ \secref{arch}), we believe this framework is able to cover a majority
of attitude-tuning needs in a simple and maintainable way, especially during
the normal operation of facilities/beamlines.  On the other hand, we are
also exploring the use of our framework during the construction phase of
facilities/beamlines, where the relations between attitude parameters and
objective parameters would often be much more complicated.  We are also aware
of requirements outside of beamlines that structurally resemble attitude
tuning, \eg\ the tuning of beams in accelerators \cite{emery2021} and the
calibration of detectors (\cf\ the \prog{xspress3-autocalib} program for the
Xspress3 readout system for silicon drift detectors and high-purity germanium
detectors); collaboration toward these directions have also been envisioned.

In the rest of this section, we would like to discuss a few issues we find
with current numerical optimisation libraries during their application in
attitude tuning.  These issues originate from physical factors that are
typically less considered in the field of numerical optimisation, and we
hope the following discussions can raise mathematicians' awareness of them.
A first of them is about the serial nature of numerical optimisation in
attitude tuning: except for tuning of simulations (\eg\ digital twins)
of physical systems, the tuning is based on manipulation of motor-like
devices and readings from detectors, which are in general not susceptible
to parallelisation.  Mathematically, this makes parallel optimisation
algorithms (genetic algorithms, particle swarm optimisation \etc)
unsuitable for attitude tuning.  Furthermore, as the efficiency of
attitude tuning often depends not only on the number of optimisation moves,
but also the trajectory length due to the movement of motors, optimisation
algorithms that aim on shortening the trajectory may be a research direction
worthy of systematic consideration.  We additionally note that fly scans
may be worth special attention in this research direction, since they can
quickly sample large numbers of points on the motion trajectories.

\begin{figure}[htbp]\centering
\includegraphics[width = 0.5\textwidth]{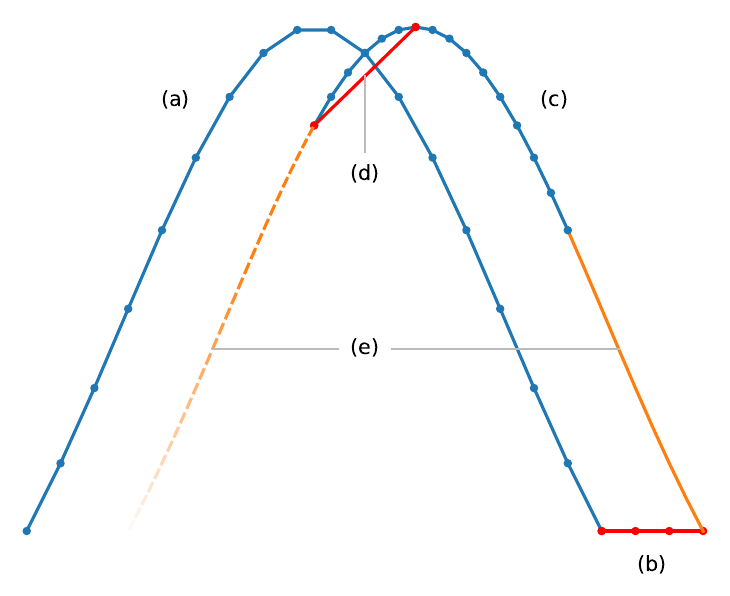}
\caption{%
	Objective parameter curve of the \texttt{max\string_parascan()} algorithm
	for each motor, where dots indicate scan points: (a) normal part of the
	coarse scan; (b) part of the coarse scan that may appear when the motor is
	stalled by a stopper; (c) the fine scan, which starts at the rightmost point
	where the objective value rised above a first threshold in the coarse scan,
	and stops when the objective value falls below a second threshold; (d) the
	final move to the peak position, deemed by the algorithm as the centre of the
	interval in the fine scan where the objective values were above the second
	threshold above; (e) intervals skipped by the fine scan to save time.%
}\label{fig:raman-hyst}
\end{figure}

Another issue we find is the precision limits in manipulation and measurement
of physical systems, \eg\ the step sizes of stepping motors; they can lead
to pathological behaviours in some algorithms on certain conditions, which
require workarounds like \verb|fix_zero()| in the file \verb|lib_4w1b.py|
(\cf\ \secref{capixes}).  On a deeper level, this is because of the assumption
in optimisation algorithms that the readback values of position $x$ were
always equal to the setpoints; similarly, most optimisation algorithms do not
consider possible measurement errors in the objective value $z$, except for
\eg\ those in the \prog{Noisyopt} library \cite{mayer2016}.  In comparison
with the minor noises in attitude parameters and objective parameters that
stem from precision limits and measurement errors, hysteresis-like effects
(\eg\ motor backlashes) and drifting of physical systems (\eg\ beam orbit
drifting in accelerators) can result in bigger troubles, and may require
special treatments in the optimisation algorithms used.  For instance,
because of multiple engineering reasons, the XRS spectrometer in
\secref{raman} has no limit switches or motor encoders, and instead only
has stopper blocks at the boundaries; so after a focusing motor gets stalled
by a stopper, the effect in \figref{raman-hyst}(b) will be observed on
the objective parameter.  Consequently, the \verb|max_parascan()| algorithm
in \secref{raman} was designed with resistance against this effect in
mind, and the resistance can be tuned with its threshold parameters.

\section{Conclusion}

The preparation steps in beamline experiments can also be of particular
interest in terms of automation, and a representative category in these
steps is attitude tuning, including beam focusing, sample alignment \etc.
We find attitude tuning a ubiquitous requirement at light sources, and a
majority of these requirements are fairly simple peak finding.  Noticing the
nature of advanced light sources and the complexity of requirements at new
light sources at HEPS, we created a versatile framework for attitude tuning
based on \prog{Mamba}.  We treat attitude tuning as a matter of numerical
optimisation, so based on the elements in numerical optimisation and
physical measurement, we implemented the \verb|AttiOptim| class which
cooperates with \prog{Bluesky}'s interfaces for motors and detectors,
as well as optimisation libraries like \verb|scipy.optimize|.  Aside from
regular peak finding, by customising \verb|AttiOptim| it is also possible to
manipulate general motor-like devices, and possible to achieve effects like
using the results from some kind of scan as the raw data for each position.
With the help from \prog{Mamba}'s infrastructure, ML/AI technologies
can also be easily integrated in our attitude tuning framework.

The first real-world example for our framework is the attitude tuning of the
polycapillary lens at 4W1B of BSRF, which demonstrates how to do simple peak
finding with straightforward processing/evaluation functions; structurally
similar attitude-tuning requirements are ubiquitous at light sources.  Also
introduced with this example is a general-purpose visualisation GUI for
attitude tuning, and the support for multi-objective tuning in our framework.
The next example is the tuning of a von Hamos XES spectrometer at 4W1B of
BSRF, which uses a much more complex evaluation function; it also shows a
way human-in-the-loop control can be integrated in attitude tuning, where
the human inputs are reused in normal ``counting'' after the tuning.  The
final example is the tuning of the XRS spectrometer at B5 of HEPS, where the
``optimisation algorithm'' interface is ``abused'' to do parallelised peak
finding of multiple objective parameters, as well as the automated assignment
of X-ray spots to analysers which is not numerical optimisation at all.

With these examples, and noticing the architectural versatility of our
framework, we believe it is able to cover a majority of attitude-tuning needs
in a simple and maintainable way.  Also reported is a virtual-beamline mechanism
based on easily customisable simulated detectors and motors, which facilitates
both testing for developers and training for users.  We noticed a few
algorithmic issues in attitude tuning, which stem from physical factors less
considered in the field of numerical optimisation, and may require attention
from mathematicians: the unsuitability of parallelisation; the importance
of shortening the trajectory of optimisation; the relevance of fly scans;
the difference between readback values and setpoints; measurement errors in
objective parameters; hysteresis-like effects and drifting of physical systems.

\section*{Acknowledgements}

The authors would like to thank the B5 beamline of HEPS and the
4W1B beamline of BSRF for a large fraction of motivations for this
paper.  This work was supported by the National Key Research and
Development Program for Young Scientists (Grant No.\ 2023YFA1609900)
and the Young Scientists Fund of the National Natural Science
Foundation of China (Grants Nos.\ 12205328, 12305371).

\bibliography{art10}
\end{document}